\documentclass[%
 reprint,
 amsmath,amssymb,
 aps,
]{revtex4-1}

\usepackage{color}

\usepackage{hyperref}
\usepackage{graphicx}
\usepackage{dcolumn}
\usepackage{bm}
\usepackage{color}

\begin{document}

	\title{Learning in brain-computer interface control evidenced by joint decomposition of brain and behavior}

\author{Jennifer Stiso$^{1,2}$}
\author{Marie-Constance Corsi$^{3,4}$}
\author{Jean M. Vettel$^{5,2,6}$}
\author{Javier Garcia$^{5,2}$}
\author{Fabio Pasqualetti$^{7}$}
\author{Fabrizio De Vico Fallani$^{3,4}$}
\author{Timothy H. Lucas$^9$}
\author{Danielle S. Bassett$^{2,8,9,10,11,12,13}$}

\affiliation{
 $^1$Neuroscience Graduate Group, Perelman School of Medicine, University of Pennsylvania,
 Philadelphia, PA 19104, USA
}
\affiliation{
$^2$Department of Bioengineering, School of Engineering \& Applied Science, University of Pennsylvania, Philadelphia, PA 19104, USA
}
\affiliation{
 $^3$Inria Paris, Aramis project-team, F-75013, Paris, France
}

\affiliation{
 $^4$Institut du Cerveau et de la Moelle Epinière, ICM, Inserm, U 1127, CNRS, UMR 7225, Sorbonne Université, F-75013, Paris, France
}
\affiliation{
 $^5$Human Research \& Engineering Directorate, US CCDC Army Research Laboratory, Aberdeen, MD, USA
}
\affiliation{
 $^6$Department of Psychological \& Brain Sciences, University of California, Santa Barbara, CA, USA
}
\affiliation{
$^7$Department of Mechanical Engineering, University of California, Riverside, CA 92521
}
\affiliation{
$^8$Department of Electrical \& Systems Engineering, School of Engineering \& Applied Science, University of Pennsylvania, Philadelphia, PA 19104, USA}
\affiliation{
$^9$Department of Neurology, Perelman School of Medicine, University of Pennsylvania, Philadelphia, PA 19104, USA
}
\affiliation{
$^10$Department of Psychiatry, Perelman School of Medicine, University of Pennsylvania, Philadelphia, PA 19104, USA
}
\affiliation{
$^{11}$Department of Physics \& Astronomy, College of Arts \& Sciences, University of Pennsylvania, Philadelphia, PA 19104, USA
}
\affiliation{
$^{12}$The Santa Fe Institute, Santa Fe, NM 87501, USA
}
\affiliation{
$^{13}$To whom correspondence should be addressed: dsb@seas.upenn.edu
}

\begin{abstract}
Motor imagery-based brain-computer interfaces (BCIs) use an individual’s ability to volitionally modulate localized brain activity, often as a therapy for motor dysfunction or to probe causal relations between brain activity and behavior. However, many individuals cannot learn to successfully modulate their brain activity, greatly limiting the efficacy of BCI for therapy and for basic scientific inquiry. Formal experiments designed to probe the nature of BCI learning have offered initial evidence that coherent activity across spatially distributed and functionally diverse cognitive systems is a hallmark of individuals who can successfully learn to control the BCI. However, little is known about how these distributed networks interact through time to support learning. Here, we address this gap in knowledge by constructing and applying a multimodal network approach to decipher brain-behavior relations in motor imagery-based brain-computer interface learning using magnetoencephalography. Specifically, we employ a minimally constrained matrix decomposition method -- \emph{non-negative matrix factorization} -- to simultaneously identify regularized, covarying subgraphs of functional connectivity, to assess their similarity to task performance, and to detect their time-varying expression. We find that learning is marked by diffuse brain-behavior relations: good learners displayed many subgraphs whose temporal expression tracked performance. Individuals also displayed marked variation in the spatial properties of subgraphs such as the connectivity between the frontal lobe and the rest of the brain, and in the temporal properties of subgraphs such as the stage of learning at which they reached maximum expression. From these observations, we posit a conceptual model in which certain subgraphs support learning by modulating brain activity in regions important for sustaining attention. To test this model, we use tools that stipulate regional dynamics on a networked system (network control theory), and find that good learners display a single subgraph whose temporal expression tracked performance and whose architecture supports easy modulation of brain regions important for attention. The nature of our contribution to the neuroscience of BCI learning is therefore both computational and theoretical; we first use a minimally-constrained, individual specific method of identifying mesoscale structure in dynamic brain activity to show how global connectivity and interactions between distributed networks supports BCI learning, and then we use a formal network model of control to lend theoretical support to the hypothesis that these identified subgraphs are well suited to modulate attention.

\end{abstract}

\maketitle

\section*{Introduction}

Both human and non-human animals can learn to volitionally modulate diverse aspects of their neural activity from the spiking of single neurons to the coherent activity of brain regions \cite{Sitaram2017, Fetz955, Sacchet2012}. Such neural modulation is made possible by routing empirical measurements of the user's neural activity to a screen or other external display device that they can directly observe \cite{Sitaram2017,Graimann2010,Moxon2015}. Referred to as a brain-computer interface (BCI), this technology can be used to causally probe the nature of specific cognitive processes \cite{Bassett2017,Reiner2014,Ros2014a,Buch2018}, and offers great promise in the treatment of neural dysfunction \cite{Shahid2010,Ros2014,Lofthouse2012,Gevensleben}. However, translating that promise into a reality has proven difficult \cite{Thibault2016,Hamedi2016,Ahn2015} due to the extensive training that is required and due to the fact that some individuals who undergo extensive training will only achieve moderate control \cite{Curran2003,Jeunet2016,Moxon2015}. A better understanding of the neural processes supporting BCI learning is an important first step towards the development of BCI therapies and the identification of specific individuals who are good candidates for treatment \cite{Curran2003,Jeunet2016}. 

While BCIs vary widely in their nature, we focus on the common motor imagery based BCIs where subjects are instructed to imagine a particular movement to modulate activity in motor cortex. Performance on motor imagery based BCIs has been associated with a diverse array of neural features, demographic factors, and behavioral measures \cite{Hammer2012,Curran2003,Bamdadian2014,Jeunet2016, Guillot2008}. Neural features predicting performance are frequently identified in areas associated with either performing or imagining action; for example, better performance is associated with higher pre-task activity in supplementary motor areas \cite{Halder2011} and larger grey matter volume in somatomotor regions \cite{Halder2011}. Interestingly, performance has also been predicted by activity in a diverse range of other cognitive systems relevant for sustained attention, perhaps due to the high cognitive demands associated with BCI learning \cite{Jeunet2016}. Specifically, better performance is associated with greater parietal power suppression in the $\alpha$ band, midline power suppression in the $\beta$ band, and frontal and occipital activation with motor power suppression in the $\gamma$ band \cite{Bamdadian2014,Grosse-Wentrup2011,Frey2013}. The role of sustained attention in BCI control is corroborated by the fact that personality and self-report measures of attention predict successful learning \cite{Hammer2012}. The heterogeneity of predictors suggests the possibility that individual differences in the interactions between cognitive systems necessary for action, action planning, and attention might explain the idiosyncratic nature of BCI control, although these interactions are challenging to quantify \cite{Fallani2018,Bassett2017}. 

Assessing the interactions between cognitive systems has historically been rather daunting, in part due to the lack of a common mathematical language in which to frame relevant hypotheses and formalize appropriate computational approaches. With the recent emergence and development of network science \cite{newman2010networks}, and its application to neural systems \cite{Bullmore2009}, many efforts have begun to link features of brain networks to BCI learning specifically and to other types of learning more generally. In this formal modeling approach \cite{Bassett2018}, network nodes represent brain regions or sensors and network edges represent statistical relations or so-called \emph{functional connections} between regional time series \cite{DeVicoFallani2014}. Recent studies have demonstrated that patterns of functional connections can provide clearer explanations of the learning process than activation alone \cite{Bassett2015}, and changes in those functional connections can track changes in behavior \cite{Bassett2017c}. During BCI tasks, functional connectivity reportedly increases within supplementary and primary motor areas \cite{Hamedi2016} and decreases between motor and higher-order association areas as performance becomes more automatic \cite{Corsi2018}. Data-driven methods to detect putative cognitive systems as modules in functional brain networks \cite{Khambhati2018,Fukushima2018} have been used to demonstrate that a particularly clear neural marker of learning is reconfiguration of the network's functional modules \cite{Bassett2017c}. Better performance is accompanied by flexible switching of brain regions between distinct modules as task demands change \cite{Bassetta,Pedersen2018,Shinea,Braun,gerraty2018dynamic}. 

While powerful, such methods for cognitive system detection are built upon an assumption that limits their conceptual relevance for the study of BCI learning. Specifically, they enforce the constraint that a brain region may only affiliate with one module at a time \cite{Khambhati2017}, in spite of the fact that many regions, comprised of heterogeneous neural populations, might participate in multiple neural processes. To address this limitation, recent efforts have begun to employ so-called \emph{soft-partitioning} methods that detect coherent patterns in mesoscale neural activity and connectivity \cite{Khambhati2017,Chai,Khambhati2016b,Chen2013,Monti2015}. Common examples of such methods are independent component analysis and principal component analysis, which impose pragmatic but not biological constraints on the orthogonality or independence of partitions. An appealing alternative is non-negative matrix factorization (NMF), which achieves a soft partition by decomposing the data into the small set of sparse, overlapping, time-varying subgraphs that can best reconstruct the original data with no requirement of orthogonality or independence \cite{Lee1999}. Previous applications of this method to neuroimaging data have demonstrated that the detected subgraphs can provide a description of time varying mesoscale activity that complements descriptions provided by more traditional approaches \cite{Khambhati2017}. For example, some subgraphs identified with NMF during the resting state have similar spatial distributions to those found with typical module detection methods, while others span between modules \cite{Khambhati2017}. As a minimally constrained method for obtaining a soft partition of neural activity, NMF is a promising candidate for revealing the time-varying neural networks that support BCI learning.

\begin{figure*}
	\begin{center}
		\centerline{\includegraphics[width=0.85\textwidth]{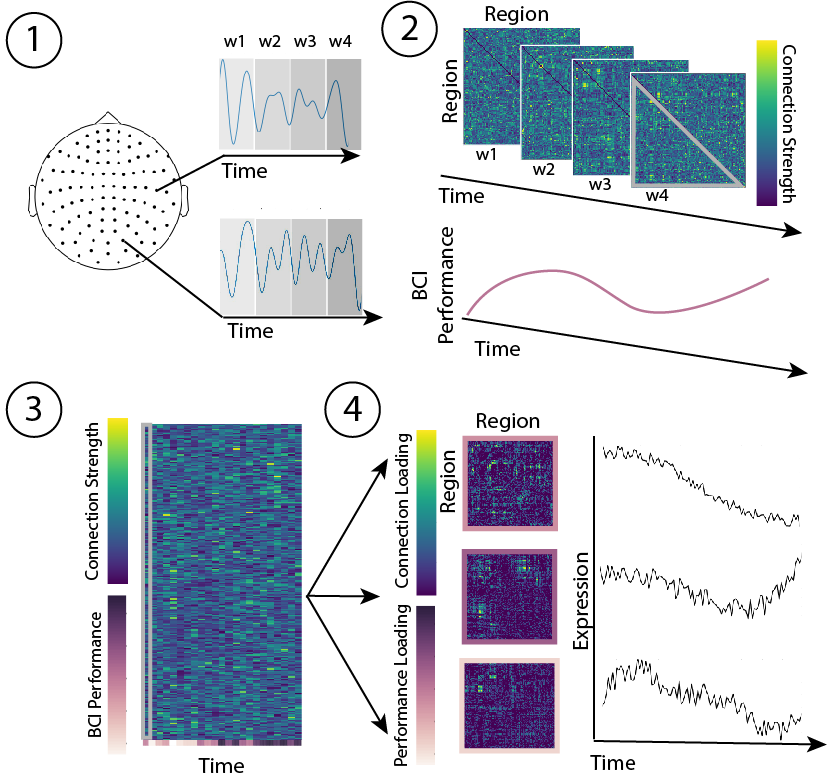}}
		\caption{\textbf{Schematic of non-negative matrix factorization.} (1) MEG data recorded from 102 gradiometers is segmented into windows ($w_{1}$, $w_{2}$, $w_{3}$, $w_{4}$, ... $w_{n}$) that each correspond to a single BCI trial. (2) In each window, functional connectivity is estimated as the weighted phase-locking index between sensor time series. The subject's performance on each trial is also recorded. (3) The lower diagonal of each trial (highlighted in grey in panel (2)) is reshaped into a vector, and vectors from all trials are concatenated to form a single data matrix. The subject's time-varying performance forms an additional row in this configuration matrix. (4) The NMF algorithm decomposes the configuration matrix (composed of neural and behavioral data) into \emph{m} subgraphs (where \emph{m} is a free parameter), with two types of information: (i) the weight of each edge in each subgraph, also referred to as the \emph{connection loading} (viridis color scale), and (ii) the time varying expression of each subgraph (black line graphs). From these data, we calculate the performance loading, or how similar the time-varying performance is to each subgraph's expression (shades of purple).}
		\label{fig:schematic}
	\end{center}
\end{figure*}

Here, we investigate the properties of dynamic functional connectivity supporting BCI learning. In individuals trained to control a BCI, we calculate single trial phase-based connectivity in magnetoencephalography (MEG) data in three frequency bands with stereotyped behavior during motor imagery: $\alpha$ (7-14 Hz), $\beta$ (15-25 Hz), and $\gamma$ (30-45 Hz). We construct multimodal brain-behavior time series of dynamic functional connectivity and performance, and apply NMF to those time series to obtain a soft partition into additive subgraphs \cite{Lee1999} (Fig.~\ref{fig:schematic}). We determine the degree to which a subgraph tracks performance by defining the \emph{performance loading} as the similarity between each subgraph's temporal expression and the time course of task accuracy. We first identify subgraphs whose performance loading predicted the rate of learning and then we explore the spatial and temporal properties of subgraphs to identify common feature across participants. We hypothesize that subgraphs predicting learning do so by being structured and situated in such a way as to easily modulate patterns of activity that support sustained attention, an important component of successful BCI control \cite{Jeunet2016}. After demonstrating the suitability of this approach for our data (Fig. S1A-B), we test this hypothesis by capitalizing on recently developed tools in network control theory, which allowed us to operationalize the network’s ability to activate regions involved in sustained attention as the energy required for network control \cite{Gu2017}. Collectively, our efforts provide a careful network-level description of neural correlates of BCI performance and learning rate, and a formal network control model that explains those descriptions.

\section*{Methods}

\subsection*{Participants}

Written informed consent was obtained from twenty healthy, right-handed subjects (aged 27.45 $\pm$ 4.01 years; 12 male), who participated in the study conducted in Paris, France. Subjects were enrolled in a longitudinal electroencephalography (EEG) based BCI training with simultaneous MEG recording over four sessions, spanning 2 weeks. All subjects were BCI-naive and none presented with medical or psychological disorders. The study was approved by the ethical committee CPP-IDF-VI of Paris.

\subsection*{BCI task}

The BCI task consisted of a standard 1 dimensional, two-target box task (14) in which the subjects modulated their $\alpha$ [8-12 Hz] and/or $\beta$ [14-29 Hz] activity to control the vertical position of a cursor moving with constant velocity from the left side of the screen to the right side of the screen. Both cursor and target were presented using the software BCI 2000 \cite{Schalk2004}. To hit the target-up, the subjects performed a sustained motor imagery of their right-hand grasping and to hit the target-down they remained at rest. Some subjects reported that they imagined grasping objects while others reported that they simply imagined clenching their hand to make a fist. Each trial lasted 7 s and consisted of a 1 s inter-stimulus interval, followed by 2 s of target presentation, 3 s of feedback, and 1 s of result presentation. In the present study, we restricted our analysis to the feedback portion of the motor imagery task because we were interested in the neural dynamics associated with learning to volitionally regulate brain activity rather than in the neural dynamics occurring at rest. BCI control features including EEG electrode and frequency were selected in a calibration phase at the beginning of each session, by instructing the subjects to perform the BCI tasks without any visual feedback. The subjects were seated in front of a screen at a distance of 90 cm. To ensure a stable position of the hands, each subject rested their arms on a comfortable support, with palms facing upward. We also recorded electromyogram (EMG) signals from the left and right arm of subjects, electrooculograms, and electrocardiograms. EMG activity was manually inspected to ensure that subjects were not moving their forearms during the recording sessions.

\subsection*{MEG Data}

\subsubsection*{Preprocessing}

As a preliminary step, temporal Signal Space Separation (tSSS) was performed using MaxFilter (Elekta Neuromag) to remove environmental noise from MEG activity. All signals were downsampled to 250 Hz and segmented into trials. ICA was used to remove blink and heartbeat artifacts. An FFT of the data from each subject was inspected for line noise, although none was found in the frequency bands studied here. We note that the frequency of the line noise (50 Hz) was outside of our frequency bands of interest. In the present study, we restricted our analyses to gradiometer sensors. Gradiometers sample from a smaller area than magnetometers, which is important for ensuring a separability of nodes as expected by network models \cite{Butts2009}. Furthermore, gradiometers are typically less susceptible to noise than magnetometers \cite{Garces}. We combined data from planar gradiometers in the voltage domain using the `sum' method from Fieldtrip's ft\_combine\_planar() function (http://www.fieldtriptoolbox.org/).

\subsubsection*{Connectivity Analysis}

To estimate phase-based connectivity, we calculated the weighted phase-locking index (wPLI) \cite{Vinck2011}. The wPLI is an estimate of the extent to which one signal consistently leads or lags another, weighted by the imaginary component of the cross-spectrum of the two signals. Using phase leads or lags allows us to take zero phase lag signals induced by volume conduction and to reduce their contribution to the connectivity estimate, thereby ensuring that estimates of coupling are not artificially inflated \cite{Vinck2011}. By weighting the metric by the imaginary component of the cross spectrum, we enhance robustness to noise \cite{Vinck2011}. Formally, the wPLI between two time series $x$ and $y$ is given by
\begin{equation}
	\phi(x,y) = \frac{|E\{imag(\Gamma_{xy})\}|}{E\{|imag(\Gamma_{xy})|\}}~,
\end{equation}

\noindent where $E\{\}$ denotes the expected value across estimates (here, centered at different samples), $\Gamma_{xy}$ denotes the cross spectrum between signals $x$ and $y$, and $imag()$ selects the imaginary component.

We first segment MEG data from gradiometers into 3 second trials, sampled at 250 Hz. The cross spectrum is then estimated using wavelet coherence \cite{Lachaux2002} in each of three frequency bands of interest ($\alpha$ 7-14 Hz, $\beta$ 15-20 Hz, and $\gamma$ 31-45 Hz), with wavelets centered on each timepoint. We chose to compute the wavelet coherence due to the fact that -- unlike Welch's method -- it does not assume stationarity of the signal \cite{Lachaux2002}. We implemented the procedure in the Fieldtrip package in MATLAB, with a packet width of 6 cycles and zero-padding up to the next power of two ('nextpow2'). We then calculate the wPLI as the mean of the imaginary component of the cross spectrum, divided by the imaginary component of the mean of the cross spectrum.

We then construct a network model of these statistical relationships where sensors ($N = 102$) are nodes, and the weight of the edge between node $i$ and node $j$ is given by the weighted phase-locking value. The graph, $G$, composed of these nodes and edges is a weighted, undirected graph that is encoded in an adjacency matrix $\mathbf{A}$. By constructing this network model, we can use statistics from graph theory and computational approaches from control theory to quantify the structure of inter-sensor functional relations \cite{Bassett2018,Bassett2017}.

\subsubsection*{Uniformly Phase Randomized Null Model}

In order to ensure that our results are not due to choices in preprocessing, the time invariant cross-correlation of neural signals, or the autocorrelation of neural signals, we repeated all of the preprocessing and analysis steps with a uniformly phase randomized null model \cite{Heitmann2018}. To enhance the simplicity and brevity of the exposition, we will also sometimes refer to this construct simply as the \emph{null} model. Surrogate data time series from the null model were calculated using a custom function in MATLAB. Essentially, the FFT of the raw data is taken, the same random phase offset is added to every channel, and then the inverse FFT is taken to return the signal to the time domain \cite{Nonlinearity}. Mathematically, this process is achieved by taking the discrete Fourier transform of a time series $y_{v}$:

\begin{equation}
Y(u)=\sum_{v=0}^{V-1} y_{v} e^{i 2 \pi u v / V},
\end{equation}

\noindent where $V$ is the length of the time series, $v$ indexes time, and $u$ indexes frequencies. We then multiply the Fourier transform by phases chosen uniformly at random before transforming back to the time domain:

\begin{equation}
\overline{y}_{v}=\frac{1}{\sqrt{V}} \sum_{v=0}^{V-1} e^{i a_{u}}\left|Y(u)\right| e^{-i 2 \pi k v / V},
\end{equation}

\noindent where the phase $a_{t} \in [0,2\pi)$.

\subsubsection*{Construction of a Multimodal Configuration Matrix}

In this work, we wished to use a data-driven matrix decomposition technique to identify time-varying subgraphs of functional connectivity that support learning. Specifically, we created a multimodal configuration matrix of edge weights and BCI performance over time, prior to submitting this matrix to a decomposition algorithm that we describe in more detail below. To construct the matrix, we first vectorize the upper triangle (not including the diagonal) of each trial's connectivity matrix, and then we concatenate all of the vectors and our one performance measure into an $E \times \tau$ matrix, where $\tau$ is the number of trials (384, if no trials were removed), and $E$ is the number of edges ($5151$) plus the number of behavioral measures ($1$). This concatenation process results in a 5152 $\times$ 384 multimodal (brain-behavior) matrix. In this task, each subject's performance is recorded as their percentage of successful trials (out of 16) on each run, and we therefore interpolate the performance time series to obtain a graded estimate of percentage correct in each trial that is $\tau$ time points long. The performance vector is then normalized to have the same mean as the other rows of the configuration matrix.

\subsection*{Non-negative Matrix Factorization}

We used a data-driven matrix decomposition method -- non-negative matrix factorization (NMF) -- to identify time-varying groups of neural interactions and behavior during BCI learning \cite{Lee1999}. Intuitively, NMF decomposes a matrix into a set of additive subgraphs with time-varying expression such that a linear combination of these subgraphs weighted by temporal expression will recreate the original matrix with minimal reconstruction error \cite{Lee1999,Khambhati2017}. The NMF algorithm can also be thought of as a basis decomposition of the original matrix, where the subgraphs are basis sets and the temporal coefficients are basis weights. Unlike other graph clustering methods \cite{Bartholomew2010,Comon2015}, NMF creates a soft partition of the original network, allowing single edges to be a part of multiple subgraphs. Additionally, unlike other basis decomposition methods \cite{Bartholomew2010,Comon2015}, NMF does not impose harsh constraints of orthogonality, or independence of the subgraphs; it simply finds the most accurate partition, given that the original matrix is non-negative. In many systems (including phase-locking) the non-negativity constraint is not difficult to satisfy, and is beneficial in physical systems where the presence of a negative weight would be difficult to interpret. 

Formally, the NMF algorithm will approximate an $E \times T$ configuration matrix $\hat{\mathbf{A}}$ by the multiplication of two matrices: $\mathbf{W}$, the subgraph matrix with dimensions $E\times m$, and $\mathbf{H}$, with dimensions $m \times T$. Here, $E$ is the number of time varying processes (behavior and functional connections derived from MEG data), $T$ is the number of time points, and $m$ is the number of subgraphs. We solve for $\mathbf{W}$ and $\mathbf{H}$ such that: 
\begin{equation}
	\underset{\mathbf{W},\mathbf{H}}{min} \frac{1}{2}||\hat{\mathbf{A}} - \mathbf{W}\mathbf{H}||^{2}_{F} + \alpha||\mathbf{W}||^2_{F} + \beta\sum_{t=1}^{T}||\mathbf{H}(:,t)||^2_1
\end{equation}

\noindent where $\beta$ is the penalty to impose sparse basis weights, and $\alpha$ is the regularization for the basis set. Regularization is frequently used in machine learning algorithms to avoid overfitting data, which is especially important when employing these techniques to examine highly variable single trial estimates of functional connectivity \cite{Kim2011a}. Additionally, selecting for sparsity will encourage the characterization of local neural processes where many edges do not contribute \cite{Khambhati2017}. From many such local processes arises the diversity of cognitive functions involved in complex tasks such as BCI control \cite{Jeunet2015}.

To solve the NMF equation, we use an alternating non-negative least squares with block-pivoting method with 100 iterations for fast and efficient factorization of large matrices, where $\mathbf{W}$ and $\mathbf{H}$ with non-negative weights are drawn from a uniform random distribution on the interval $[0, 1]$\cite{Kim2014}. The parameter $m$ is drawn from the range (2,20), and $\alpha$ and $\beta$ are drawn from the range (0.001,2). We select for parameters that will both minimize the residual error, and maximize the temporal and subgraph sparsity \cite{Khambhati2017}. Specifically, we select the optimal parameters $\bar{m}$, $\bar{\alpha}$, and $\bar{\beta}$ that are in the lowest 25$_{th}$ percentile for residual error, and the highest 25$_{th}$ percentile for temporal and subgraph sparsity. This procedure resulted in an average $\bar{m}$ of 7.4, an average $\bar{\alpha}$ of 0.46, and an average $\bar{\beta}$ of 0.45. Distributions of parameters and reliability across runs are shown in Fig. S2 and S3. 

Given the non-deterministic nature of this approach, we also test for the stability of our identified clusters using a consensus clustering algorithm \cite{Greene2008}. Our procedure was comprised of the following ordered steps: (1) run the NMF algorithm $r$ = 100 times per multimodal configuration matrix, (2) concatenate the subgraph matrix $\mathbf{W}$ across $r$ runs into an aggregate matrix with dimensions $E \times (r ∗ \bar{m})$, and (3) apply NMF to the aggregate matrix to determine a final set of subgraphs and expression coefficients \cite{Khambhati2017}. While the implementation is heuristic in nature, we found that across two runs of the algorithm, we obtain highly consistent selections for parameters (see Supplement), bolstering confidence in the robustness of the subsequent analyses.

\subsubsection*{Subgraph Inclusion}
Most subgraphs are sparse, with distributions of temporal coefficients skewed towards zero (see Fig. S4). However, for every subject and every frequency band, one subgraph showed very little regularization (no edges were equal to 0) and had a uniform, rather than skewed distribution of temporal coefficients. These subgraphs are clear outliers from the others, and appear to be capturing global phase-locking across the entire brain, rather than any unique subsystem. To answers questions about the time varying interactions between neural systems, we were interested in differences between the subgraphs that were spatially localized, having edges regularized to zero. Because including these outlier subgraphs would obscure those differences, we removed these subgraphs from all further analyses.

\begin{figure}
	\begin{center}
		\centerline{\includegraphics[width=0.5\textwidth]{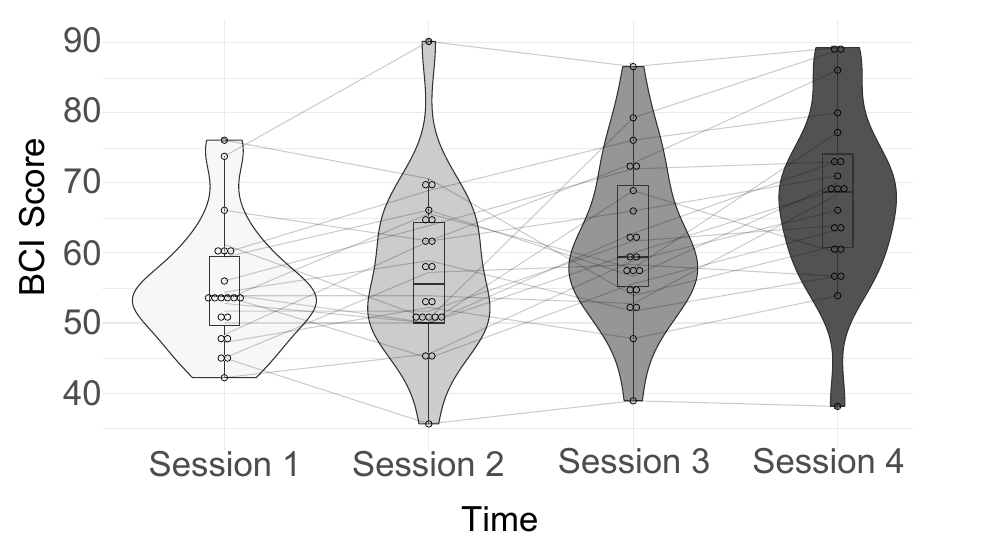}}
		\caption{\textbf{BCI performance.} Each subject's average performance across four days within two weeks. BCI Score is the percentage of correct trials during that session.}
		\label{fig:behavior}
	\end{center}
\end{figure}

\subsection*{Group Average Subgraphs}

After applying NMF to the multimodal brain-behavior matrix, we next turned to a study of the nature of the detected subgraphs after ranking them by performance loading. Specifically, we were initially interested in determining which edges contributed to each ranked subgraph most consistently across the population. For this purpose, we used a consistency based approach to create a group representative subgraph for each ranked subgraph \cite{Roberts2017}. In this procedure, each subject's subgraph was first thresholded to retain only the 25\% strongest connections (see Fig. S5 for evidence that results are robust to variations in this choice). We then constructed an average $N\times N$ subgraph $G$, where $N$ is the number of channels and where each element $G_{ij}$ quantifies how many subjects (out of 20) displayed an edge between region $i$ and region $j$ in their thresholded subgraph. In addition to visually depicting these group representative subgraphs, we also wished to summarize their content in large anatomical regions. We therefore binned edges into 10 anatomically defined areas in frontal, motor, parietal, occipital, and temporal lobes in both hemispheres. Lobes were obtained from BrainStorm \cite{Tadel:2011} software (neuroimage.usc.edu/brainstorm/Tutorials/MontageEditor) (Fig. S8).

\subsection*{Optimal Control}

Our final broad goal was to provide a theoretical explanation for why certain networks support BCI learning. We hypothesized that these regularized networks might have structures that make it easier for the brain to modulate the patterns of activity that are necessary for BCI control. This hypothesis motivated us to formulate and validate a model to explain how the sparse statistical relationships characteristic of each subgraph could support the production of brain activity patterns implicated in BCI learning \cite{Gu2016,Betzel2016}. Additionally, this model should account for the brain's ability to reach these patterns of activity in the context of the BCI task, where there is increased volitional modulation of the left motor cortex. Here, we use tools from network control theory to satisfy these conditions \cite{Pasqualetti2014a}. Specifically, we characterize the theoretical brain activity at each sensor as a vector $x(t)$, and we use the adjacency matrix $\mathbf{A}$ of a subgraph to quantify the ease with which that activity can affect other regions. We then incorporate volitional input control as input into the brain ($u(t)$) at a specific region (given by $\mathbf{B}$). Then, by stipulating

\begin{equation}
\dot{x}(t) = \mathbf{A}x(t) + \mathbf{B}u(t),
\end{equation} 

\noindent we model the linear spread of activity along the connections in $\mathbf{A}$ in the context of input to regions given in $\mathbf{B}$.

With this model of network dynamics, optimal control trajectories can be formalized and identified by developing a cost function that seeks to minimize two terms: (i) the distance of the current state from the target state and (ii) the energy required for control. Specifically, we solve the following minimization problem:

\begin{equation}
\begin{split}
\underset{\mathbf{u}}{min}\int_{0}^{T}(\mathbf{x}_{T} - \mathbf{x}(t))^{T}(\mathbf{x}_{T} - \mathbf{x}(t)) + \rho\mathbf{u}_{\kappa}(t)^{T}\mathbf{u}_\kappa dt, \\
s.t. \quad\dot{\mathbf{x}} = \mathbf{Ax}(t) + \mathbf{Bu}(t),\quad\mathbf{x}(0) = \mathbf{x}_{0},\quad and\quad\mathbf{x}(T) = \mathbf{x}_{T} ,
\end{split}
\end{equation}

\noindent where $\rho$ is a free parameter that weights the input constraint, $\textbf{x}_T$ is the target state, and $T$ is the control horizon, which is a free parameter that defines the finite amount of time given to reach the target state. During BCI control, there is specific, targeted control to a specific area of the brain (here, the left motor cortex) in addition to other ongoing control and sensory processes. We wished for our selection of the input matrix $\mathbf{B}$ to reflect this richness and also allow for computationally tractable calculations of optimal control, which is difficult for sparse control sets. Therefore, we constructed the input matrix $\mathbf{B}$ so as to allow input that was dominated by the BCI control site, while maintaining minor contributions from other areas. More specifically, rather than being characterized by binary state values, channels other than the one located over left motor cortex were given a value of approximately $5\times10^{-5}$ at their corresponding diagonal entry in $\mathbf{B}$. See Supplement for the full derivation from \cite{Gu2016}. 

It is important to note that in general the tools from linear controllability theory are not applicable to the functional networks commonly derived from neuroimaging data for two reasons. The first reason is that the model which the tools are built upon stipulates a time-dependent propagation of activity along edges; such a propagation is physically true for structural connections derived from white matter, but is not generally true for other types of connections used in network models, such as morphometric similarity or most common functional connectivity measures. While we do not expect that these simple models will fully capture neural dynamics, it is important to explore how much variance these models do explain, even if we expect that amount to be small. The second reason is that the model assumes that interactions between nodes "a" and "c" are not due to node "b", an assumption that is violated by measures of statistical similarity such as the Pearson correlation coefficient which is the measure of functional connectivity most commonly employed in neuroimaging studies. Because we are using neither structural connectivity nor common measures of functional connectivity, it was necessary for us to first prove that the networks we are studying are consistent with our model. To address the first point regarding the propagation of activity along edges, we demonstrate that the structure of the subgraphs used have utility in predicting empirical brain state transitions, and that the relative contribution of each subgraph is related to its temporal expression (Fig. S1C-D). It is only in light of these validations that we are able to interpret our results as a potential model for driving brain activity. To address the second point regarding isolation of pairwise relations not due to third party effects, we note that the matrix $\mathbf{A}$ that we study reflects statistical similarity in phase after strict regularization that removes redundant statistical relationships (Fig. S1A-B).

\subsubsection*{Target state definition}

A central hypothesis in this work is that certain regularized subgraphs are better suited to drive the brain to patterns of activity that are beneficial for BCI control than others. To test this hypothesis, we create target states that reflect these beneficial patterns, based on previous literature. Target states for motor imagery and attention are obtained for each band individually from references \cite{Bamdadian2014,Grosse-Wentrup2011,Frey2013}, and can be briefly described as follows: $\alpha$ contralateral motor suppression for motor imagery and parietal suppression for attention, $\beta$ contralateral motor suppression and ipsilateral motor activation for motor imagery and vertex suppression for attention, and $\gamma$ contralateral motor activation for motor imagery and motor cortex suppression with frontal and occipital activation for attention (Fig. S9). Channels were divided into lobes using standard montages provided by Brainstorm \cite{Tadel:2011} software (neuroimage.usc.edu/brainstorm/Tutorials/MontageEditor). The target state of channels in brain regions where we did not have specific hypotheses for their activity were set to zero; the target state of channels with activation were set to 1 and that of channels with deactivation were set to -1. Initial states were set to 0 for all channels. We then calculate the optimal energy (using the optimal control equation described above) required to reach each of these target states to test the hypothesis that subgraphs that support learning will have lower energy requirements than those that do not.

\subsection*{Statistical Analyses}

Much of our analyses involve testing differences in distributions across subjects for different subgraphs or sessions, both for phase-randomized and empirical data. We also compare these distributions to subject learning rate defined as the slope of performance over time. For the results displayed in Fig. 2 here in the main manuscript, we used a repeated measures ANOVA  to test for the presence of a main effect across conditions given that the distributions of performances were normal (see Fig. S10). In Fig. 3 here in the main manuscript, we sought to predict learning rate with ranked performance loading. After plotting quantile-quantile plots (see Fig. S11-S13) for the learning rate, and each of the performance loadings, it became clear that the lowest loadings were not normally distributed. Therefore, we used a linear model combined with non-parametric testing utilizing 5000 permutations (lmPerm package in R https://cran.r-project.org/web/packages/lmPerm). Standardized coefficients were calculated using the lm.beta package in R (https://cran.r-project.org/web/packages/lm.beta/lm.beta.pdf). We use a Bonferroni correction to control false positive errors due to multiple comparisons across all 6 predictors ($\alpha$ = 0.008). To obtain an estimate of how sensitive our results are to our specific sample, we also plot summary statistics from 500 models obtained from bootstrapping a sample of equal size ($N = 60$, 3 band and 20 subjects). To examine differences in consistency (Fig. 4 here in the main manuscript), we use a linear model ($consistency \sim band + dataType + rank$) to test for a main effect of data type (null or empirical), band, and subgraph on consistency (see Fig. S14). We next sought to determine if different subgraphs had consistently different temporal expression for null and empirical data (Fig. 5 here in the main manuscript). We also used a repeated measures ANOVA to test for a main effect of subgraph across bands, and paired $t$-tests to test for differences amongst individual subgraphs (Fig. S15). Lastly, for the results shown in Fig. 6 here in the main manuscript, we test the relationship between learning rate and optimal control energy differences for several different models. Pearson's correlations were used, given that the data appears normally distributed and has few outliers (see Fig. S16-S19).

\subsection*{Data and Code}
Code for analyses unique to this manuscript are available at github.com/jastiso/netBCI. Code for the NMF algorithm and the NMF parameter selection is available at github.com/akhambhati/Echobase/tree/master/Echobase
/Network/Partitioning/Subgraph. Code for optimal control analyses is available at github.com/jastiso/NetworkControl. Data necessary to reproduce each figure will be made available upon request.

\section*{Results}

\subsection*{BCI Learning Performance}

Broadly, our goal was to examine the properties of dynamic functional connectivity during BCI learning, and to offer a theoretical explanation for why a certain pattern of connectivity would support individual differences in learning performance. We hypothesized that decomposing dynamic functional connectivity into additive $N \times N$ subgraphs would reveal unique networks that are well suited to drive the brain to patterns of activity associated with successful BCI control. We use MEG data from 20 healthy adult individuals who learned to control a motor-imagery based BCI over four separate sessions spanning a two week period. Consistent with prior reports of this experiment \cite{Corsi2018}, we find a significant improvement in performance across the four sessions (one-way ANOVA $F(3,57) = 13.8$, $p = 6.8\time10^{-7}$) (Fig.~\ref{fig:behavior}). At the conclusion of training, subjects reached a mean performance of 68\%, which is above chance (approximately 55 - 60\%) level for this task \cite{Muller-Putz2008}.

\subsection*{Dynamic patterns of functional connectivity supporting performance}

To better understand the neural basis of learning performance, we detected and studied the accompanying patterns of dynamic functional connectivity. First, we calculated single trial phase-based connectivity in MEG data in three frequency bands: $\alpha$ (7-14 Hz), $\beta$ (15-25 Hz), and $\gamma$ (30-45 Hz). We then used non-negative matrix factorization (NMF) -- a matrix decomposition method -- to separate the time-varying functional connectivity into a soft partition of additive subgraphs. We found that the selected parameters led to an average of 7.4 subgraphs, with a range of 6 to 9, and that all frequency bands had a decomposition error lower than 0.47 (mean $\alpha$ error = 3.52, mean $\beta$ error = 0.379, mean $\gamma$ error = 0.465) (Fig. S2). The error is the Frobenius norm of the squared difference between our observed and estimated connectivity matrices (with dimensions $5152 \times 384$) and takes values between 0 and 1. For each band, the error value is low, giving us confidence that we have fairly accurately reconstructed relevant neural dynamics. To determine whether any properties of the identified subgraphs were trivially due to preprocessing choices, NMF parameters, or time-invariant autocorrelation in neural activity, we repeated the full decomposition process after permuting the phases of all time series uniformly at random. We found that the statistics of subgraph number and decomposition error were similar for the uniformly phase randomized data, indicating that any differences in subgraph and temporal expression between null and empirical data is not due to the NMF algorithm's inability to find a good decomposition, but rather due to the structure of the chosen decomposition (Fig. S2).

\begin{figure}
	\begin{center}
		\centerline{\includegraphics[width=0.48\textwidth]{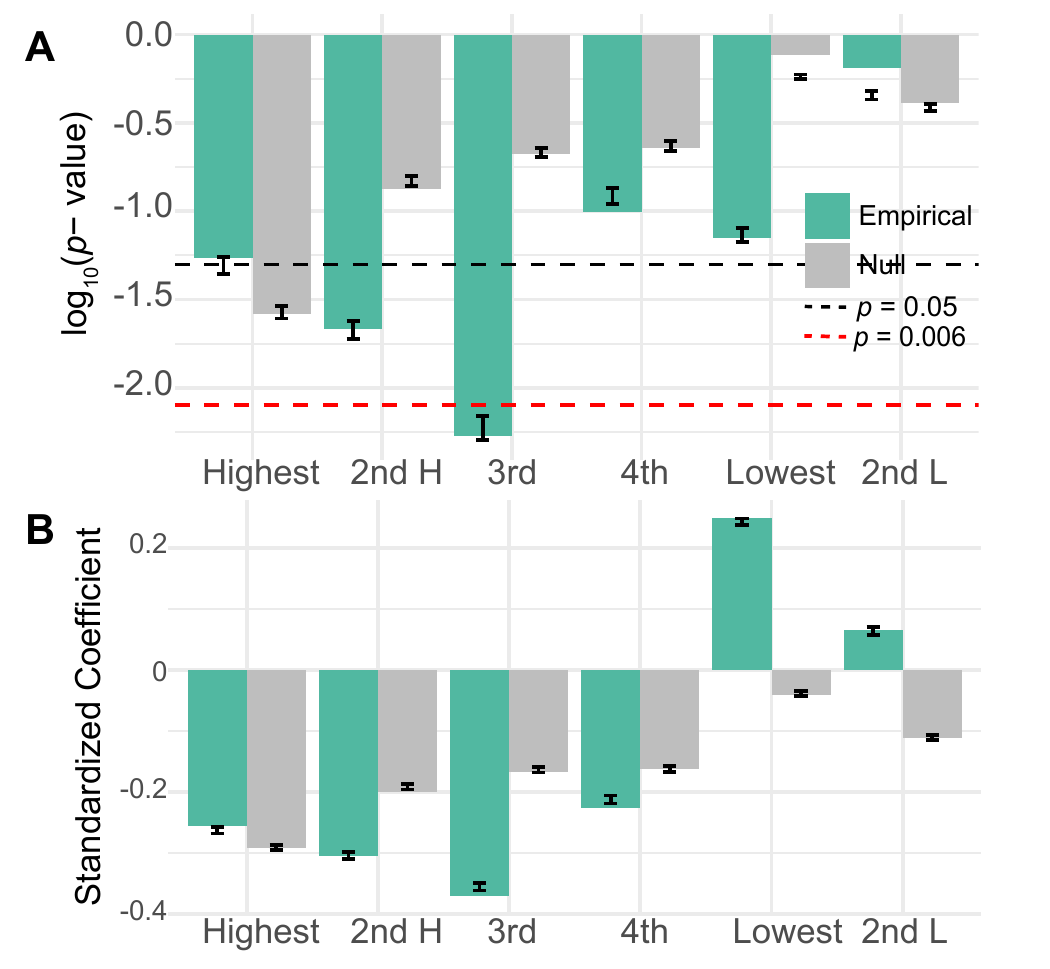}}
		\caption{\textbf{Performance loading predicts learning.} \emph{(A)} Here we show the $p$-values for empirical (green) and uniformly phase randomized (grey) data for linear models predicting the slope of performance with ranked performance loading from each frequency band. The black line corresponds to $p = 0.05$, while the red dashed line corresponds to the Bonferroni corrected $\alpha = 0.008$. Error bars show the standard error and median of $p$ values from 500 models with bootstrapped samples. \emph{(B)} The standardized regression coefficients for the same models. Error bars show the standard error and mean of coefficients from 500 models with bootstrapped samples.}
		\label{fig:loading}
	\end{center}
\end{figure}

We quantified the similarity between each subgraph's temporal expression and the time course of performance, and we refer to this quantity as the subgraph's performance loading (Fig.~\ref{fig:schematic}). We hypothesized that the ranked performance loading would predict task learning, as operationalized by the slope of performance over time. It is important to note the distinction between performance and learning: performance is defined as task accuracy and therefore varies over time, while learning is defined as the linear rate of change in that performance over the course of the experiment (384 trials over 4 days).  We tested whether learning was correlated with the performance loading of subgraphs. Because the minimum number of subgraphs in a given subject was 6, we decided to investigate the top four highest performance loading subgraphs, and the smallest and second smallest nonzero loading subgraphs. We found a general trend that the performance loading from high loading subgraphs was negatively associated with learning rate, and the performance loading from low loading subgraphs was positively associated with learning rate (Fig.~\ref{fig:loading}AB). We assessed the statistical significance of these trends and found that only the third highest loading subgraph displayed a performance loading that was significantly correlated with learning rate after Bonferroni correction for multiple comparisons (linear model with permutation tests $slope \sim loading3 + band:$ $p = 0.005$). Performance loading from uniformly phase randomized surrogate data for this subgraph did not predict learning rate ($p = 0.292$). The direction of the observed effect in the empirical data is notable; subjects with \emph{lower} loading onto \emph{high} loading subgraphs learned the task better, suggesting that learning is facilitated by a dynamic interplay between several subnetworks. It is also notable that the highest loading subgraphs are not the strongest predictors of learning, indicating that the subgraphs that most closely track performance are not the same as the subgraphs that track changes in performance.

\subsection*{Spatial properties of dynamic patterns of functional connectivity}

Next we sought to better understand why the third highest loading subgraph most robustly predicted learning. We hypothesized that because of this subgraph's predictive power across subjects, it might recruit consistent brain regions and reflect the involvement of specific cognitive systems across subjects. To evaluate this hypothesis, we began by investigating the shared spatial properties of this subgraph in comparison to the others. To identify shared spatial features we grouped subgraphs together by their ranked performance loading, and then quantified how consistent edges were across participants \cite{Roberts2017} (see Methods). We found that the average consistency varied by frequency band, and differed between the empirical and surrogate data, but not across ranked subgraphs (linear model $consistency \sim band + rank + data:$ $F_{band}(2,17) = 90.36$, $p_{band} = 9.00\times10^{-10}$, $F_{data}(1,17) = 41.8$, $p_{data} = 5.78\times10^{-6}$). The $\alpha$ band had the most consistent edges, followed by the $\gamma$ band, and then the $\beta$ band ($t_{\alpha\beta} = -12.68$, $p_{\alpha\beta} = 4.3\times10^{-10}$, $t_{\alpha\gamma} = -10.41$, $p_{\alpha\gamma} = 1.2\times10^{-8}$). In the uniformly phase randomized surrogate data, we observed less consistent subgraphs than those observed in the empirical data ($t = -6.47$, $p = 5.78\times10^{-6}$). These observations support the conclusion that across the population, despite heterogeneous performance, similar regions interact to support performance and learning to varying degrees.

\begin{figure*}
	\begin{center}
		\centerline{\includegraphics[width=0.9\textwidth]{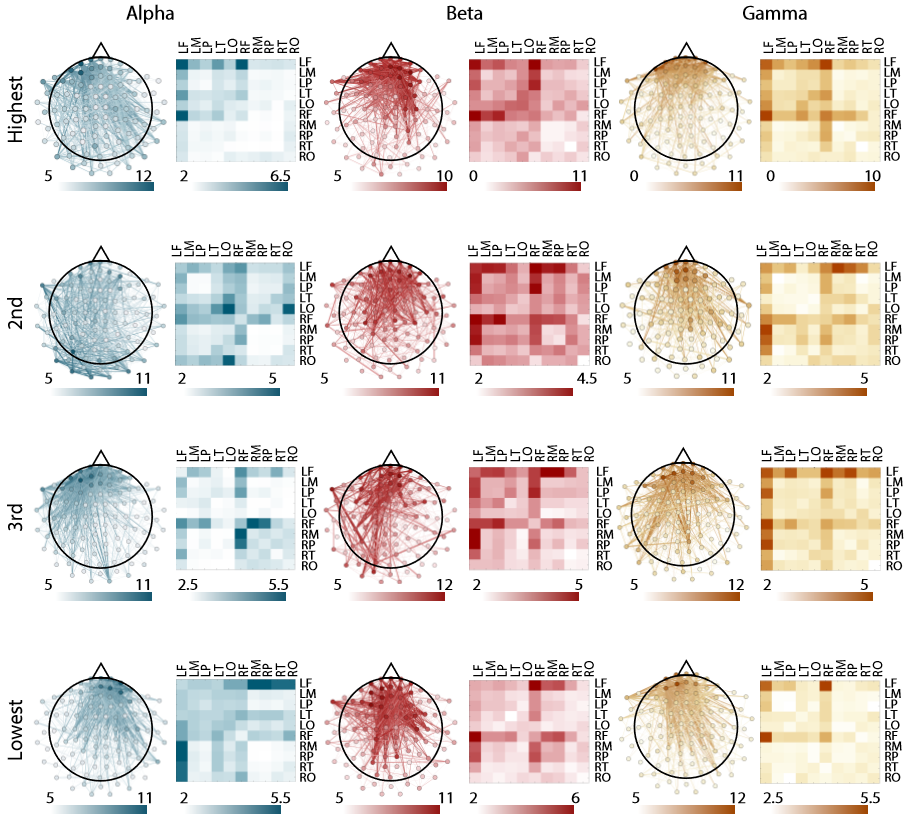}}
		\caption{\textbf{Spatial distribution of subgraph edges that are consistent across participants.} Consistent edges for each frequency band and for each ranked subgraph. Left images show individual edges plotted on a topographical map of the brain. Right images show the mean edge weight over sensors for a given region. We studied 10 regions, including the frontal lobe, temporal lobe, parietal lobe, occipital lobe, and motor cortex in both hemispheres. The weight of the edge corresponds to the number of individual participants for whom the edge was among the 25\% strongest for that subgraph.}
		\label{fig:spatial}
	\end{center}
\end{figure*}

Anatomically, subgraphs were dominated by connectivity in the frontal lobe sensors, with subtle differences in the pattern of connections from the frontal lobe sensors to sensors located in other areas of the brain (Fig.~\ref{fig:spatial}). To determine which functional edges were most consistent in each subgraph and frequency band, we calculated the average consistency over each lobe and motor cortex in both hemispheres (for the same analysis in surrogate data, see Fig. S6). In the $\alpha$ band, the most consistent edges on average were located in the left frontal lobe in the highest performance loading subgraph, in the left occipital lobe in the second highest performance loading subgraph, between right frontal and right motor in the third highest performance loading subgraph, and between left frontal lobe and right parietal lobe in the lowest performance loading subgraph. In the $\beta$ band, the most consistent edges were located between right and left frontal lobe for the highest and second highest performance loading subgraph, between left frontal lobe and right motor for the third highest performance loading subgraph, and between left and right frontal lobe for the lowest performance loading subgraph. In the $\gamma$ band, the most consistent edges were located in the left frontal and right frontal lobes for the highest performance loading subgraph, in the left frontal lobe and right motor for the second highest performance loading subgraph, and in left frontal and right frontal lobe for the third highest and lowest performance loading subgraphs. We also note that the most consistent individual edges for each subgraph are still only present in 10-12 individuals, indicating a high amount of individual variability. Collectively, these observations suggest widespread individual variability in the spatial composition of ranked subgraphs, with the most consistent connectivity being located in the frontal lobe during BCI learning. 
 
\subsection*{Temporal properties of dynamic patterns of functional connectivity}
  
Importantly, subgraphs can be characterized not only by their spatial properties, but also by their temporal expression. We therefore next examined the temporal properties of each subgraph to better understand why the third highest performance loading subgraph most robustly predicted learning. As a summary marker of temporal expression, we calculated the total energy of the time series operationalized as the sum of squared values, as well as the time of the peak value of the time series. Across frequency bands, we found no significant dependence between energy and subgraph ranking. We did find a significant effect of rank for the peak time of temporal expression obtained from the empirical data (repeated measures ANOVA $peak \sim rank + band:$ $F_{rank}(3,215) = 6.67$, $p_{rank} = 2.53\times10^{-4}$ but not from the uniformly phase randomized surrogate data ($F_{rank}(3,215) = 1.28$, $p = 0.282$). Overall, peak times are widely distributed across individuals. However we find that across bands, the highest performance loading subgraph has a later peak, which is intuitive since performance is generally increasing over time and these subgraphs most strongly track performance. 

\begin{figure}
 	\begin{center}
 		\centerline{\includegraphics[width=0.5\textwidth]{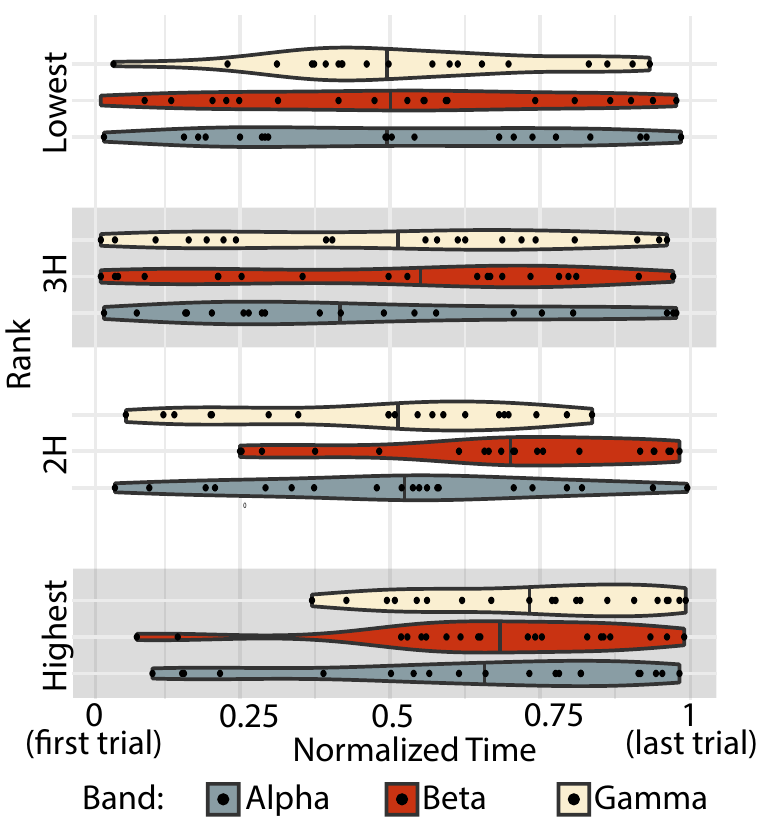}}
 		\caption{\textbf{Temporal expression of ranked subgraphs.} The peak temporal expression for every subject (black data point), for each frequency band (indicated by color) and for each subgraph (ordered vertically). Violin plots show the density distribution of all subjects' peaks. The median is marked with a solid line through the violin plot.}
 		\label{fig:expression}
 	\end{center}
 \end{figure}

We then performed \emph{post-hoc} paired $t$-tests corrected for multiple comparisons (Bonferroni correction $\alpha$ = 0.006) between the highest performance loading subgraph and all other ranked subgraphs in each band. In the $\alpha$ band, the highest performance loading subgraph only peaked significantly later than the lowest (paired $t$-test $N$ = 20, $t_{low} = 8.06$, $p_{low} = 1.49\times10^{-7}$) after Bonferroni correction ($\alpha$ = 0.006). In the $\beta$ band, the highest performance loading subgraph peaked significantly later than all others (paired $t$-test $N = 20$, $t_{2H} = 10.9$, $p_{2H} = 1.39\times10^{-9}$; $t_{3H} = 7.56$, $p_{3H} = 3.57\times10^{-7}$; $t_{low} = 8.07$, $p_{low} = 1.49\time10^{-7}$). In the $\gamma$ band, the highest performance loading subgraph peaked significantly later than the second highest, and lowest loading subgraphs (paired $t$-test $N$ = 20, $t_{2H} = 4.50$, $p_{2H} = 2.46\times10^{-4}$; $t_{low} = 8.06$, $p_{low} = 1.49\times10^{-7}$). (Fig.~\ref{fig:expression}). Finally, we asked whether the time of the peak in the third highest performance loading subgraph predicted learning. We did not find a relationship between peak time and learning in any frequency band (Pearson's correlation: $\alpha:$ $r = 0.005$, $p = 0.98$, $\beta:$ $r = 0.047$, $p = 0.84$, $\gamma: r = -0.21$, $p = 0.037$). To summarize these findings, we note that across participants and especially in the $\beta$ band, subgraphs that support performance are highly expressed late in learning, when performance tends to be highest. However, subgraphs that support learning do not have consistent peaks across subjects, and each individual's peak does not predict their learning rate, indicating that some other feature of these subgraphs must explain their role in learning.

\subsection*{Explaining dynamic patterns of functional connectivity supporting BCI learning via network control theory}
 
Lastly we asked how the third highest loading subgraph could facilitate successful BCI performance, as shown in Fig.~\ref{fig:loading}. Here, we considered an edge -- extracted under penalties of spatial and temporal sparsity -- as a potential path for a brain region to affect a change in the activity of another brain region \cite{Fox2012,Fox2013,Ferreri2014}. Assuming the true connectivity structure is sparse, the regularization applied in the NMF algorithm can remove large statistical relationships between regions that are not directly connected, but might receive common input from a third region \cite{Shen2016,Das2017} (see Methods for addition discussion, and see Fig. S1A-B for the effect of regularization on the prevalence of triangles). We hypothesized that the pattern of edges in this subgraph would facilitate brain states, or patterns of activity, that were predictive of BCI literacy. Specifically, we expected that when the brain mirrored the connectivity of the third subgraph, the brain could more easily reach states of sustained motor imagery or sustained attention than when the brain mirrored the connectivity of the lowest performance loading subgraph. We also hypothesized that the magnitude of this difference would be predictive of each subject's learning rate. To test these hypotheses, we used mathematical models from network control theory to quantitatively estimate the ease with which the brain can reach a desired pattern of activity given a pattern of connectivity (see Methods and Fig. S1C-D for analyses demonstrating the efficacy of the regularized subgraphs in linearly predicting changes in activity). Specifically we calculated the optimal control energy required to reach a target state (either sustained motor imagery or sustained attention) from an initial state when input is applied primarily to the left motor cortex, which was the site of BCI control (Fig. 6A-B). 
 
\begin{figure*}
 	\begin{center}
 		\centerline{\includegraphics[width=0.95\textwidth]{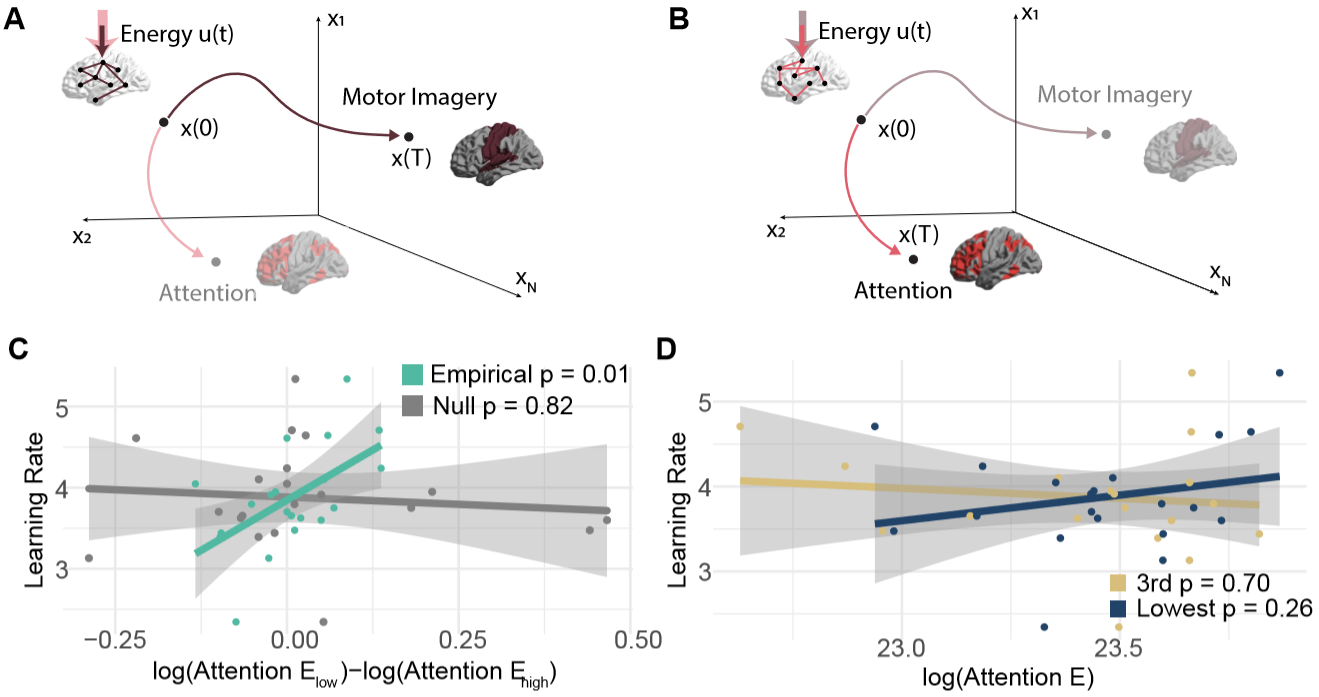}}
 		\caption{\textbf{Separation of the ability to modulate attention predicts learning.} Different patterns of connections will facilitate transitions to different patterns of brain activity. We hypothesize that the ease with which connections in certain regularized subgraphs facilitate transitions to patterns of activity that support either motor imagery \emph{(A)} or attention \emph{(B)} will predict learning rate. We use network control theory to test this hypothesis. We model how much energy ($u(t)$) is required to navigate through state space from some initial pattern of activity $x(0)$ to a final pattern of activity $x(T)$. Some networks (e.g., the brown network in panel \emph{A}) will require very little energy (schematized here with a smaller, solid colored arrow) to reach patterns that support motor imagery, while other networks (e.g., the pink network in panel \emph{B}) will have small energy requirement to reach patterns of activity that support attention. \emph{(C)} The relationship between learning rate and the difference in energy required to reach the attention state when the underlying network takes the form of the lowest versus third highest performance loading subgraphs for empirical data (green) and uniformly phase randomized surrogate data (grey). \emph{(D)} The relationship between the learning rate and the energy required to reach the attention state when the underlying network takes the form of the lowest performance loading subgraph, or when the underlying network takes the form of the third highest performance loading subgraph.}
 		\label{fig:control}
 	\end{center}
\end{figure*}

We tested whether the third highest performance loading subgraph supported the transition to states of sustained motor imagery or sustained attention with smaller energy requirements than other subgraphs that did not support learning in the same way. We chose the lowest performance loading subgraph for comparison because it was the only subgraph with a large positive standardized regression coefficient for predicting learning, which contrasts sharply with the large negative coefficient for the third subgraph. For both states (motor imagery and attention), we found no population level differences in energy requirements by the two subgraphs (paired $t$-test $N = 20$, motor imagery: $t_{\alpha} = -0.005$, $p_{\alpha} = 0.565$, $t_{\beta} = 1.38$, $p_{\beta} = 0.184$, $t_{\gamma} = -1.00$, $p_{\gamma} = 0.329$. attention: $t_{\alpha} = -1.35$, $p_{\alpha} = 0.193$, $t_{\beta} = -0.344$, $p_{\beta} = 0.735$, $t_{\gamma} = -0.937$, $p_{\gamma} = 0.360$). We next tested whether the magnitude of the difference in energy required by the two subgraphs to reach a given state tracked with learning rate. In the $\beta$ band, we observed a significant correlation between the magnitude of the energy difference to reach attentional states and learning rate over subjects (Pearson's correlation coefficient $r = 0.560$, $p = 0.0103$, Bonferroni corrected for multiple comparisons across frequency bands; Fig.~\ref{fig:control}). Notably, the relationship remained significant when controlling for subgraph density (linear model $slope \sim energy\_difference + density\_difference$: $t_{energy} = 2.68$, $p_{energy} = 0.0158$, $t_{density} = -0.266$, $p_{density} = 0.794$). When using subgraphs derived from the uniformly phase randomized surrogate data, the relationship was not observed (Pearson's correlation $r = -0.0568$, $p = 0.819$). We next asked which subgraph contributed most to this effect. We found no significant relationship between learning rate and the energy required to reach the attentional state by the third highest performance loading subgraph (Pearson's correlation $r = -0.389$, $p = 0.702$) or by the lowest performance loading subgraph (Pearson's correlation $r = 0.227$, $p = 0.335$). This finding suggests that learning rate depends on the relative differences between subgraphs, rather than the energy conserving architecture of one alone. As a final test of specificity, we assessed whether this difference was selective to the third highest and lowest performance loading subgraph. We found no significant relationship when testing the difference of the highest with the third highest performance loading subgraph (Pearson's correlation $r = -0.554$, $p = 0.586$), the highest with the lowest performance loading subgraph (Pearson's correlation $r = 0.40$, $p = 0.077$), the second highest with the third highest performance loading subgraph (Pearson's correlation $r = 0.266$, $p = 0.257$), or the second highest with the lowest performance loading subgraph (Pearson's correlation $r = -0.072$, $p = 0.764$). This pattern of null results underscores the specificity of our finding.
 
\subsection*{Reliability and specificity of inferences from network control theory}
 
Collectively, our findings are consistent with the hypothesis that during BCI learning, one subnetwork of neural activity arises, separates from other ongoing processes, and facilitates sustained attention. An alternative hypothesis is that our results are due to trivial factors related to the magnitude of the attentional state, or could have just as easily been found if we had placed input to a randomly chosen region of the brain, rather than to the left motor cortex which was the actual site of the BCI control. To determine whether these less interesting factors could explain our results, we performed the same network control calculation but with a spatially non-overlapping target state, and then -- in a separate simulation -- with a mirrored input region (right motor cortex rather than left motor cortex). We performed the spatial shifting by ordering the nodes anatomically (to preserve spatial contiguity), and then circular shifting the attention target state by random number between $1$ and $N - 1$.  For 500 circularly shifted states, only 3 (0.6\%) had a correlation value equal to or stronger than the one observed (Fig. S7). Furthermore, we found no significant relationship between learning rate and the difference in energy required by the two subgraphs to reach the true attention state when input was applied to the right motor cortex instead of the left motor cortex (Pearson's correlation $t = 0.711$, $p = 0.313$). Together, these two findings suggest that the relationship identified is specific to BCI control.

Finally, we assessed the robustness of our results to choices in modeling parameters. First we performed the computational modeling with two different sets of control parameter values (see Supplement). In both cases, the significant relationship remained between learning rate and the difference in energy required by the two subgraphs to reach the attentional state (set one Pearson's correlation coefficient $r = 0.476$, $p = 0.0338$; set two Pearson's correlation coefficient $r = 0.514$, $p = 0.0204$). Second, since our target states were defined from prior literature, there was some flexibility in stipulating features of those states. To ensure that our results were not unduly influenced by these choices, we tested whether ideologically similar states would provide similar results. Namely, we assessed (i) the impact of varying the magnitude of (de)activation by changing (-)1 to (-)2, (ii) the impact of the neutral state by changing 0 to 1, and (iii) the impact of negative states by changing -1, 0 and 1 to 1, 2, and 3. We found a consistent relationship between learning rate and the difference in energy required by the two subgraphs to reach the attentional state when we changed the magnitude of activation/deactivation (Pearson's correlation coefficient $r = 0.560$, $p = 0.0103$), as well as when we changed the neutral state (Pearson's correlation coefficient $r = 0.520$, $p = 0.0188$). However, we found no significant relationship when removing negative states (Pearson's correlation coefficient $r = 0.350$, $p = 0.130$), indicating that this result is dependent on our choice to operationalize deactivation as a negative state value. After performing these robustness checks, we conclude that a selective separation of the third highest and lowest performance loading subgraphs impacts their ability to drive the brain to patterns of sustained attention in the $\beta$ band in the context of BCI control. This result is robust to most of our parameter choices, is selective for biologically observed states, and is not observed in surrogate data.

\section*{Discussion}

In this work, we use a minimally constrained decomposition of dynamic functional connectivity during BCI learning to investigate which groups of phase locked brain regions (subgraphs) support BCI control. The performance loading onto these subgraphs favors the theory that dynamic involvement of several subgraphs during learning supports successful control, rather than extremely strong expression of a single subgraph. Additionally, we find a unique role for the third highest loading subgraph in predicting learning at the population level. This result shows that learning is not simply predicted by the subset of edges that has the most similar temporal expression, but rather that a subnetwork with a middling range of similarity has the strongest relationship with performance improvement. While the spatiotemporal distribution of this subgraph was variable across individuals, we did observe some consistencies at the group level. Spatially, the third highest loading subgraph showed strong edges between left frontal and right motor cortices for low frequencies, and left frontal and left motor cortices for the $\gamma$ band. Lower frequencies showed stronger connectivity to the ipsilateral (to imagined movement) motor cortex, suggesting a possible role in suppression for selective control. This subgraph also showed the highest expression earlier than the other ranked subgraphs we investigated, perhaps linking it to the transition from volitional to automatic control.

We next wished to posit a theory of how these subgraphs fit with previously identified neural processes important for learning, despite their heterogeneity across subjects. After quantifying the extent to which NMF regularization removed potentially redundant relationships between regions (Fig. S1A-B), we suggested that the regularized pattern of statistical relationships identified in this subgraph could comprise an avenue through which brain activity could be modulated via cognitive control or external input. We then hypothesized that these networks would be better suited to modulate activity in either regions implicated in attention or in motor imagery than other subgraphs, and further that individuals whose networks better modulated activity in these regions would display greater task learning \cite{Jeunet2016}. We chose to operationalize the ``ease of modulation'' with a metric from network control theory called \emph{optimal control energy}. Optimal control energy quantifies the minimum input needed to drive the brain from an initial pattern of activity to a final pattern of activity, while also assuring that the pattern of activity stays close to the target state at every point in time. This last constraint assures that we do not pass through biologically unfeasible patterns of activity to reach our desired pattern. The notion of optimal control energy that we use here assumes a particular linear model of how neural dynamics change given potential avenues of communication between regions. Importantly, in the supplement (Fig. S1C-D) we show that our subgraphs predict empirical brain state changes according to this model, and that the contribution of each subgraph to empirical changes in brain state is related to its temporal expression. Using this model, we did not find any population differences in optimal control energy when the simulation was enacted on the third highest performance loading subgraph compared to the lowest performance loading subgraph. However, we did find that the magnitude of this difference predicted learning in individual subjects. This result was specific to the $\beta$ band and to brain regions implicated in attention. Critically, the relation to learning could not be explained by the energy of either subgraph alone, was not present in surrogate data derived from a uniformly phase randomize null model, and was robust to parameter choices. Overall, the observations support our hypothesis that in the $\beta$ band the subgraphs we identified that support learning are well suited to modulate activity in brain regions associated with attention. 

\subsubsection*{A delicate balance of interactions is required for BCI learning}

Our initial analysis explored the relationship between performance loading and learning. It is important to note the behavioral difference between performance and learning: we use the term performance to refer to task accuracy over time, whereas we use the term learning to refer to how well a subject is able to increase that accuracy. With that distinction in mind, we aimed to better understand how subgraphs that vary similarly to performance (those with high performance loading) relate to learning. We found that the subgraph with the third highest performance loading best predicted learning and that a narrow distribution of performance loading across all subgraphs was associated with better learning. Together, these two observations are in line with previous research in motor and spatial learning, which shows that some brain structures display differential activity during learning that is independent of performance \cite{Shelton2004,Purushotham2002}. Our work adds to this literature by demonstrating that in addition to targeted differences in individual brain regions or networks, a minimally constrained decomposition of dynamic functional connectivity across the whole brain reveals that separable processes are most associated with performance and with learning. 

Additionally, we find that BCI learning is not predicted simply by the processes most strongly associated with performance and learning individually, but by a distributed loading across many different subgraphs. This notion is supported by the modestly predictive role of standard deviation of loading predicting learning, and also by the sign of beta value predictors for ranked subgraphs. Generally, subgraphs with higher ranked loading were negative predictors, while subgraphs with lower ranked loading were positive predictors. A wealth of whole brain connectivity analyses have similarly shown that the interaction between systems is an important component of skill learning specifically, and other domains of learning more generally \cite{Bassett2015,Blainey2014,Mantzaris2013,Bassett2017c}. While we observed marked interactions between many regions, the majority were located in the frontal lobe for all frequency bands. Previous work has also demonstrated changes in frontal-motor \cite{Ma2009,Leff2016} and fronto-parietal \cite{JaniceLin2013} connectivity during motor skill learning. In BCI learning specifically, the strength of white matter connectivity between frontal and occipital regions predicts control of motor imagery based BCIs \cite{Sitaram2013}. Additionally, analyses of this same experiment have shown task related changes in functional connectivity were spatially diffuse, and found in frontal, temporal, and occipital regions in the $\alpha$ band \cite{Corsi2018}, and were strongest in frontal, motor, central, and parietal regions in the $\beta$ band. Our results add to these findings by demonstrating that the most consistent regions that covary in their functional connectivity are interactions between the frontal lobe and other regions. Our work shows that broad motifs like the dynamic integration of multiple systems (including cognitive systems involving the frontal lobe) found in other types of learning are also important for BCI learning. Additionally, we add to previous work on BCI learning specifically by quantifying the structure of covarying subgraphs of connectivity.

\subsubsection*{BCI learning is heterogenous across individuals}

We find population level consistencies in spatial and temporal properties of ranked subgraphs despite having no constraint to assure consistency across individuals. However, we also note that there is a high degree of variability in both of these measures. The variability is mirrored in the subjects' performance, with final performances varying from 38.1 \% to 89.3 \%. Our observations are in line with previous literature demonstrating variability in subjects' performance and learning for psychological, cognitive, and neurological predictors \cite{Jeunet2016, Halder2010}. Such pervasive and marked individual differences presents a challenge for the use of BCIs clinically \cite{Brunner2010}. To address this challenge, researchers have explored ways to optimize BCI features and algorithms for neurofeedback itself \cite{Vidaurre2011,Kubler2015} and to identify selection criteria for BCI based therapies \cite{Jeunet2015,Halder2010}. The results of our study support the idea that different individuals will have slightly different neural correlates of both performance and learning based on a variety of features such as demographics \cite{Schumacher}, spatial manipulation skills \cite{Vuckovic2013}, relationship with the technology \cite{Brosnan1998}, and attention span \cite{Grosse-Wentrup2011,Grosse-Wentrup2012}. Our findings also highlight the importance of studying models fit to each individual when searching for selection criteria for BCI therapies. Here, despite temporal and edge level heterogeneity, our minimally constrained, individual specific method of brain connectivity decomposition revealed a robust predictor of learning with a theoretical role that aligns well with previous literature. Further development and expansion of this model to incorporate resting state neuroimaging data and other physiological predictors could be a promising direction for selection of candidates for BCI therapies before training.

\subsubsection*{Role of beta oscillations in BCI learning}

Prominent theories describing the neural processes that give rise to cognition and shape our behavior often involve integration of complex multimodal information using a combination of top-down predictions (built from prior experience) and bottom–up, sensory-driven representations of the dynamic world around us \cite{Doehrmann2008, Kok2012, Talsma2015}. These generalized frameworks, in turn, require the precise coordination of ensemble neural activity both within and between brain regions. Several theoretical approaches have examined how these two scales of functional activity may harmonize to produce the desired behavior \cite{Bressler2001}, and empirical research has shown that there is consistent cross-talk between these scales \cite{Buzsaki2004}. Within human neuroimaging work, synchronous oscillations have been critical to understanding this complex coordination, where cortico-cortical propagation delays and membrane potentials give rise to observed oscillatory activity in the brain \cite{Bastos2012,Singer2001}. Here, we study the time varying connectivity within $\alpha$, $\beta$, and $\gamma$ bands. Much like how specialized functions arise from different brain regions, different narrowband oscillations have been implicated in diverse but specialized processes, where some generalizable theories suggest a role for $\alpha$ in disengagement of task irrelevant areas or a lack of sensory processing \cite{Palva2007}, $\beta$ in sustaining the current cognitive state \cite{Engel2010} and $\gamma$ in task active local cortical computation \cite{Fries2009}. Specifically in the context of motor imagery based BCIs, $\alpha$ and $\beta$ bands have prominent signatures in motor imagery \cite{McFarland2000}. Our results show that only the $\beta$ band’s functional connectivity is well suited to modulate patterns of activity that support sustained attention (not motor imagery), which is a critical process for BCI control. While our results are in line with generalized theories on the role of oscillations in cognition, the specificity of the $\beta$ band in our results extends classic studies that discuss the role of this oscillation in attention \cite{PfurtschellerAb1997} and in maintaining the current cognitive state \cite{Engel2010}. Our results suggest that this maintenance, a consistent control (or attention to) internally generated activity, may play a crucial role in longterm BCI use.

\subsubsection*{Methodological Considerations}
\emph{NMF}
Non-negative matrix factorization is a machine learning technique for separating, in our case, a multimodal configuration matrix into a soft-partition of subgraphs with time-varying expression. This process has several advantages, such as being able to link behavioral and neural data, and creating a quantification of mesoscale structure where brain regions can participate in multiple functional groups. Nevertheless, the method also faces several limitations that are common to other large-scale machine learning techniques. NMF yields a low rank approximation of a large configuration matrix, and can sometimes be rank deficient for large number of subgraphs, for very large datasets, or for datasets with high covariance. Because of this sensitivity, we were not able to test our data against independently phase randomized null models.

\emph{Spatial Resolution}
We chose to complete our analyses in sensor, rather than source space. This choice limits the anatomical resolution of our data, and therefore the specificity of the claims that we can make about the spatial distribution of the regions involved. However, source reconstruction requires many parameter choices and has potentially confounding effects on estimates of functional connectivity \cite{Brookes2011a,Hillebrand2012,Colclough2016}. Additionally, we were not interested in finer resolution distribution of the identified subgraphs, but more in the process of identifying them, in validating the hypothesis that features of these subgraphs predicted learning, and in their theoretical functions.

\emph{Optimal Control}
We chose to use tools from network control theory to quantify the ease with which each network can modulate brain activity. Network control theory relies on several assumptions that should be considered when interpreting these results. First, the model of dynamics that we employ is linear and noise free, unlike the brain \cite{Gu2014}, but has proven useful in gaining intuitions about the behavior of nonlinear systems \cite{Muldoon2016,Honey}. However, we still sought to quantify the ability of this linear model to explain empirical changes in brain state. Specifically, we asked two questions: (1) do the regularized subgraphs used in our analyses have the ability to predict state transitions, and do they do so better than randomly rewired networks, and (2) is the contribution of each subgraph to explaining a given state transition proportional to its temporal expression, and is it more proportional than a different subgraph's temporal expression? To evaluate these questions, we generated brain states for every trial (band specific power at each channel) and simulated Eq. 5 (see Supplement). Regarding the similarity of predicted and empirical state transitions, we find modest correlation values (mean Pearson's $r = 0.25$) that are significantly greater than the correlations observed from randomized networks. Similarly for our second question, we found small but positive correlations between the contribution of each subgraph to a given transition and its temporal expression (mean Pearson's $r = 0.03$), which was also significantly greater than correlations to temporal expression from mismatched subgraphs. While it is unsurprising that our linear model did not fully capture neural dynamics across a three second trial, it is worth considering extensions that can maximize this similarity for future analyses investigating how connections between regions facilitate changes to activity. One option is to use effective connectivity -- such as autoregressive models \cite{Shen2016,Neumaier} -- that solve for a network of connections that best predicts the evolution of brain states in time. However, effective connectivity matrices are often sparse, and therefore not well suited to NMF matrix decomposition used in the present work.  Alternatively, one could use non-linear models of dynamics \cite{Jirsa1996} and non-linear control theory \cite{Zanudo2016} to capture a wider range of dynamic behaviors, although non-linear control does not currently support the same scope of tools available for linear control theory. Lastly, future work could use functional approximation\cite{Brunton2016} in order to identify a set of simple basis functions that well approximate the data. If a sparse approximation can be found, it supports the idea that the underlying non-linear dynamics can be captured with linear combinations of these basis functions, and therefore are suitable to be modeled with simplified linear models.

Additionally, network control is typically applied to time invariant, structural connections that have a clear role as an avenue along which brain activity can propagate. Here we used functional connectivity (weighted phase locking) which is a statistical relationship that (1) does not imply the presence of a physical connection and (2) is not time invariant. Due to (1), our original functional connectivity matrix can have large values between two regions that are not directly connected, but might both connect to the same region. This situation would lead to a triangle composed of three connections in a functional connectivity matrix where in reality there are only two connections. However, the regularization applied by the NMF algorithm mitigates this concern in a manner that is similar to the regularization applied in effective connectivity metrics \cite{Shen2016,Das2017}. We also explicitly quantify the effect of regularization on triangles in our subgraphs and find a dramatic reduction from the original functional connectivity (Fig. S1A-B). This quantification, along with the two validations discussed above, show that our model is a suitable way to evaluate the role of regularized subgraphs in modulating different patterns of activity. In relation to (2), we note that functional connectivity in not time-invariant, unlike the state matrix more commonly employed in linear control models. However, it is important to note that NMF identifies subgraphs that are separable from their temporal expression, and that we expect that the hypothesized role in control would only be prominent when the subgraph was highly expressed.

\subsubsection*{Conclusion}
In conclusion, we use a minimally constrained method of matrix decomposition that is specific to each human participant to investigate the dynamic neural networks that support BCI learning. We find that the subgraphs that most tightly mirror performance are not the same subgraphs that most strongly support learning. Additionally, we find that the interaction between many different neural processes is important for BCI learning. While the subgraphs identified are heterogeneous (as is subject performance), we find consistent involvement of frontal and motor cortices in subgraphs that support learning. We also observe differential temporal expression amongst subgraphs, and perhaps most notably that the subgraphs that vary more similarly with performance reach their highest expression later in learning. Lastly, we test the hypothesis that subgraphs that support learning are better suited to modulate activity in brain regions important for attention than other subgraphs. We find evidence to support this hypothesis in the $\beta$ band specifically, ultimately suggesting that the separation of processes for maintaining attention is important for successful BCI learning. Our results align with prior work from dynamic functional connectivity in other types of skill learning, and also highlight a method for identifying individual predictors of successful BCI control with theoretical support.

\section*{Acknowledgements}
We would like to thank Ankit N. Khambhati for helpful discussions regarding the application of NMF to BCI learning, and Marcelo G. Mattar for helpful discussions regarding joint decomposition of brain and behavior. We would also thank Pragya Srivastava, Jason Kim, Lia Papadopoulos, and Eli Cornblath for their helpful comments on the manuscript. D.S.B. and J.S. acknowledge support from the John D. and Catherine T. MacArthur Foundation, the Alfred P. Sloan Foundation, the ISI Foundation, the Paul Allen Foundation, the Army Research Laboratory (W911NF-10-2-0022), the Army Research Office (Bassett-W911NF-14-1-0679, Grafton-W911NF-16-1-0474, DCIST- W911NF-17-2-0181), the Office of Naval Research, the National Institute of Mental Health (2-R01-DC-009209-11, R01 – MH112847, R01-MH107235, R21-M MH-106799), the National Institute of Child Health and Human Development (1R01HD086888-01), National Institute of Neurological Disorders and Stroke (R01 NS099348), the National Science Foundation (BCS-1441502, BCS-1430087, NSF PHY-1554488 and BCS-1631550), and French program ”Investissements d’avenir” ANR-10-IAIHU-06; ”ANRNIH CRCNS” ANR-15-NEUC-0006-02. The content is solely the responsibility of the authors and does not necessarily represent the official views of any of the funding agencies.
4

\bibliography{mendeley}

\end{document}